\documentclass[fleqn,twoside]{article}
\usepackage{espcrc2}
\usepackage{amssymb}
\usepackage{psfig}

\newcommand{\beq}{\begin{equation}}
\newcommand{\eeq}{\end{equation}}
\newcommand{\beqn}{\begin{eqnarray}}
\newcommand{\eeqn}{\end{eqnarray}}
\newcommand{\bea}[1]{\beq\begin{array}{#1}}
\newcommand{\eea}{\end{array}\eeq}

\newcommand{\tr}{\mathop{\rm Tr}}

\newcommand{\ket}[1]{|\,#1\,\rangle}
\newcommand{\bra}[1]{\langle\,#1\,|}

\newcommand{\diff}{\partial}
\newcommand{\cC}{{\cal C}}

\newcommand{\PL}[3]{{\it Phys. Lett. }{\bf #1} (#2) #3}

\newcommand{\PR}[3]{{\it Phys. Rev. }{\bf #1} (#2) #3}
\newcommand{\JL}[3]{{\it JETP Lett. }{\bf #1} (#2) #3}
\newcommand{\CMP}[3]{{\it Comm. Math. Phys. }{\bf #1} (#2) #3}

\title{
\thispagestyle{empty}
\vspace{-25mm}
\rightline{\small ITEP-LAT/2003-22~~~~~}
\rightline{\small KANAZAWA-03-28~~~~~}
\vspace{10mm}
On the non-Abelian Stokes theorem in SU(2) gluodynamics\thanks{Presented by F.V.G. at Lattice'03.}
}
\author{F.V.~Gubarev
  \address{ITEP, B.Cheremushkinskaya 25, Moscow, 117259, Russia}
  \address{Institute for Theoretical Physics, Kanazawa University, Kanazawa 920-1192, Japan}
  \thanks{F.V.G. is supported by JSPS Fellowship P03024.}
}
\begin{document}

\begin{abstract}
We discuss the non-Abelian Stokes theorem for SU(2) gauge fields
which avoids both additional integration variables and surface ordering.
The idea is to introduce the instant color orientation of the flux
piercing the loop. The non-Abelian Stokes theorem is also considered
on the lattice and various terms contributing to the trace of the
Wilson loop are discussed.
\end{abstract}

\maketitle

Nowadays, there are quite a few
formulations of non-Abelian Stokes theorem (NAST) available
(for review see, e.g., Ref.~\cite{Broda:2000id} and references therein).
Generically there exist two principal approaches, the
operator~\cite{Halpern:1978ik} and the path-integral~\cite{Diakonov:fc} one.
We consider a new version of non-Abelian Stokes theorem \cite{self}
for $SU(2)$ valued Wilson loops in the fundamental
representation. The basic idea is to introduce the instant color orientation
of the chromo-magnetic flux piercing the loop, which
allows us to avoid the path ordering
and represent the Wilson loop phase as an ordinary integral to which
Abelian Stokes theorem applies.

Consider Wilson loop $W(T)$ in the fundamental representation
evaluated on a closed contour $\cC = \{ x_\mu(t), t\in [0;T], x_\mu(0) = x_\mu(T)\}$,
parameterized by $x_\mu(t)$. By definition $W(T)$ provides a solution to
the following Schr\"odinger like equation
\beq
\label{defEQ}
\bra{\psi(t)}\, \left( i \diff_t ~+~ A \right) ~=~ 0\,,
\eeq
\beq
\label{prop}
\bra{\psi(t)}=\bra{\psi(0)}W(t) =
\bra{\psi(0)}\,\mathrm{T}e^{i \int_0^t A(\tau) d\tau}\,,
\eeq
where $A$ is the tangential component of the gauge potential
and $\bra{\psi}$ is a vector in the spin-1/2 irreducible representation space (IRR)
of $SU(2)$ group. Thus the Wilson loop $W(t)$ can be interpreted
as a quantum mechanical evolution operator with the time-dependent Hamiltonian
$H = - A(t)$. The corresponding state space coincides with spin-1/2 IRR,
in which a convenient basis  is provided by generalized
(spin) coherent states~\cite{Perelomov:1971bd}
(see, e.g., Ref.~\cite{Perelomov} for review).
The spin coherent states $\{\bra{\vec{n}}\}$ are parameterized
by a set of unit three-dimensional vectors $\vec{n}$, $\vec{n}^2 = 1$ and
in this basis arbitrary state $\bra{\psi}$ has a unique representation
\beq
\label{coherentRep}
\bra{\psi} ~=~ e^{ i\varphi} \,\, \bra{\vec{n}}\,,
\quad
\bra{\vec{n}}\,\,g ~=~ e^{ i\varphi(g,\vec{n})} \,\, \bra{\vec{n}_g}\,.
\eeq
Therefore, Eq.~(\ref{prop}) becomes
\beq
\label{constant}
\bra{\psi(t)} ~=~ e^{ i\varphi(t)} \,\, \bra{\vec{n}(t)} ~=~ \bra{\vec{n}(0)} \, W(t)\,,
\eeq
where without loss of generality we have taken $\varphi(0) = 0$.

Eq.~(\ref{defEQ}) imposes no restrictions on the initial vector $\bra{\vec{n}(0)}$.
However, there exists a distinguished initial state
which is of particular importance for the discussion below. Namely, let us take
$\bra{\vec{n}(0)}$ to be the eigenstate of the full evolution operator
\beq
\label{eigen}
\bra{\vec{n}(0)}\,W(T) = e^{ i\varphi(T)} \bra{\vec{n}(0)}\,.
\eeq
Note that generically $1/2\tr W(T)=\cos\varphi(T)\ne\pm 1$ and we assume this from now on.
{}From Eqs.~(\ref{defEQ},\ref{coherentRep}) is follows that~\cite{Gubarev:2000qg}
\beq
\label{Phase0}
\varphi(T)=\int_0^T \Big(
\bra{\vec{n}}A\ket{\vec{n}} - i \bra{\vec{n}}\diff_t\ket{\vec{n}} \Big)dt\,.
\eeq
Using standard properties of the spin coherent
states~\cite{Perelomov} one can represent (\ref{Phase0}) in vector-like notations
\bea{c}
\label{Phase}
1/2\tr W(T)= \\
\rule[-4mm]{0mm}{9mm}
=\cos[ \frac{1}{2}\int_C \vec{n}\vec{A}\,dt+\frac{1}{4} \int_{S_C}
\vec{n} \cdot [ \partial\vec{n} \times \partial\vec{n}]
\,d^2\sigma] = \\
=\cos[\frac{1}{4} \int_{S_C}\{\vec{n}\vec{F}_{\mu\nu}+
\vec{n}\cdot [ D\vec{n} \times D\vec{n}]\}\,
d^2\sigma]\,,
\eea
where $\vec{n}(t)\in \cC$ has been smoothly extended
to $\vec{n}(\sigma)\!\in\!S_\cC$,
$D^{ab}_\mu = \delta^{ab}\diff_\mu - \varepsilon^{acb} A^c_\mu$
is the covariant derivative and
$F^a_{\mu\nu} = \diff_\mu A^a_\nu - \diff_\nu A^a_\mu -  \varepsilon^{abc} A^b_\mu A^c_\nu$
is the non-Abelian field strength.

By construction the state $\bra{\vec{n}(t)}$ is an eigenstate of $W^+(t)W(T)W(t)$.
Thus $\vec{n}(t)$ is the instant color orientation of the flux piercing the loop $\cC$.
Since $\cC$ is not infinitesimal the color direction of the flux varies with $t$.

The assignment of the vector field $\vec{n}(t)$ to a given closed path
$\cC$ is unique only up to the sign since Eq.~(\ref{eigen})
possesses two solutions
\beq
\label{eigen2}
\bra{\,\pm\,\vec{n}(0)}\,\, W(T) ~=~ e^{\,\pm\,i\varphi(T)} \,\,
\bra{\,\pm\,\vec{n}(0)}\,.
\eeq
However, this sign ambiguity is only global: if $\bra{\vec{n}(0)}$
is an eigenstate with $\varphi(T) > 0$, then for any $t\in[0;T]$ $\bra{\vec{n}(t)}$
is an eigenstate of $W^+(t) \, W(T) \, W(t)$ with the same positive phase. Therefore there is
only a global freedom to change $\vec{n}(t)\to - \vec{n}(t)$ for all $t$ simultaneously.

It is amusing to note that Eq.~(\ref{Phase}) looks similar
to the non-Abelian Stokes theorem of Ref.~\cite{Diakonov:fc}.
We conclude therefore that the path integral of Ref.~\cite{Diakonov:fc}
is exactly saturated by the two particular trajectories $\pm\,\vec{n}(t)$. It is worth
mentioning however that Eq.~(\ref{Phase}) does not correspond in general
to any semi-classical approximation.

Notice that Eq.~(\ref{Phase}) is still not uniquely defined since
the vector field $\vec{n}(t)$ may be extended arbitrarily from $\cC$ to $S_\cC$.
The only requirement is that the extension 
$\cC \ni \vec{n}(t) \to \vec{n}(\sigma)\in S_\cC$ must be continuous and
the distribution $\vec{n}(\sigma)$ must agree with $\vec{n}(t)$
at the boundary $\delta S_\cC = \cC$. On the other hand, Eq.~(\ref{Phase})
is applicable to any infinitesimal area element of $S_\cC$
and therefore naturally defines the direction of $\vec{n}(\sigma)$. Then the only problem
is the choice of sign of $\vec{n}(\sigma)$. Here the continuity requirement 
becomes crucial. Indeed, since the surface $S_\cC$ is assumed to be smooth
the field $\vec{F}\delta\sigma$ is continuous on $S_\cC$ and hence
$\vec{n}(\sigma)$ being the eigenstate of $1+i\vec{\sigma}\vec{F}\delta\sigma$
might be defined continuously as well.

The derivation of the non-Abelian Stokes theorem on the lattice
may found in Ref.~\cite{self}. The final result
\beq
\label{NAST-3}
\varphi_w ~=~ \sum\limits_{x\in S_\cC} \varphi_x ~+~ 
\sum\limits_{x\in S_\cC} \Omega_x ~+~  \sum\limits_{x\in\cC} \alpha_x
\eeq
\centerline{\psfig{file=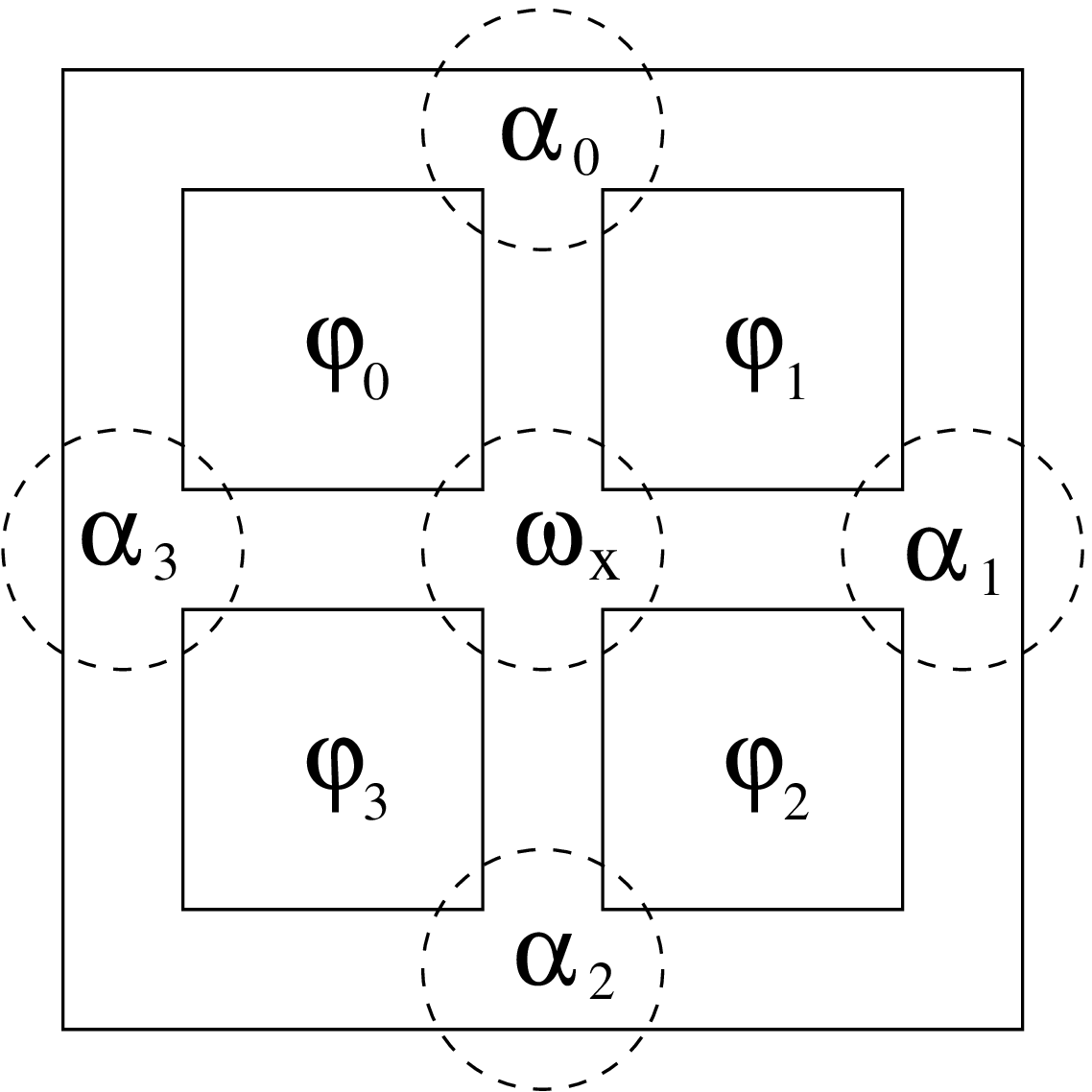,width=0.2\textwidth,silent=}}
\centerline{Fig.~1}

~

\noindent
is illustrated on Fig.~1. It is understood that phases $\varphi_w$,
$\varphi_x$ are calculated via Eq.~(\ref{eigen}). $\varphi_w$ is the phase
of the large Wilson loop, $1/2\tr W(\cC) = \cos\varphi_w$, where $\cC$ is the
planar $2\times 2$ closed contour bounding the surface $S_\cC$.
The first term on the r.h.s. ('dynamical part') is the sum of contributions coming
from four 'internal' plaquettes belonging to $S_\cC$. In particular,
$1/2\tr U_{p_x}=\cos\varphi_x$, $p_x= 0,...,3$,
where $U_{p_x}$ is the corresponding plaquette matrix. The second term ('solid angle')
comes from the points common to four different 'internal' plaquettes.
By construction there are four
color vectors per plaquette situated at plaquette's corners.
Then $\Omega_x$ is the oriented solid angle between four vectors at point $x$.
The third term ('perimeter contribution') is analogous
to the second one. It accounts for the difference in color direction between
the states on the nearest to the loop 'internal' plaquettes and the states on the loop
itself.

Eq.~(\ref{NAST-3}) has a simple physical interpretation. The magnitude of the total flux,
$\varphi_w$, piercing large closed contour $\cC$ is the sum of several terms.
The first term sums up the magnitudes of elementary fluxes penetrating the surface
$S_\cC$. Since the theory is non-Abelian each elementary flux has its own color
orientation which is no less important than the flux magnitude.
The other terms in Eq.~(\ref{NAST-3}) take into account the difference in color
orientation of various fluxes on $S_\cC$ as well as of the total flux piercing $\cC$. 

One can show~\cite{self} that in the limit $a\to 0$ Eq.~(\ref{NAST-3}) formally
reproduces Eq.~(\ref{Phase}). Moreover, the 'dynamical part' and 'solid angle'
correspond to two terms under cosine in Eq.~(\ref{Phase}), while the 'perimeter'
contribution provides correct boundary conditions.
However, this conclusion is only valid for $\vec{n}_x$, $x\in S_\cC$
continuous across the plaquette boundaries. This suggests a natural way
to fix the relative sign of eigenstates on neighboring plaquettes
analogously to the continuum considerations. Namely, we propose to fix
the particular distribution of eigenstates by the requirement that
\beq
\label{min}
R ~=~ \sum\limits_{x\in S_\cC} |\Omega_x| ~+~
\sum\limits_{x\in\cC} |\alpha_x|
\eeq
takes the minimal possible value (it is assumed, of course, that eigenvectors
at the boundary $\vec{n}(t) \in \cC$ are held fixed from the very beginning).
This prescription fixes completely and unambiguously all the states
$\vec{n}(\sigma)\in S_\cC$  provided that the functional $R$ has a unique minimum.
The uniqueness of the minimum of $R$ is a separate issue and we have
no analytical methods to investigate it. However, at least
numerically the minimum of (\ref{min}) might be approximated with high accuracy.

We performed simple  lattice experiments with Eq.~(\ref{NAST-3})
in pure SU(2) lattice gauge theory on $12^4$ lattice at $\beta=2.4$
using the standard Wilson action. Since the decomposition (\ref{NAST-3})
is gauge invariant it is legitimate to ask what are the expectation values
\bea{ccc}
\label{exp}
\rule[-2mm]{0mm}{0mm}
\langle\exp\{\,\, i \sum_{x\in S_\cC} \varphi_x\,\,\}\rangle & \sim & e^{-T V_{dyn}(R)}\,,\\
\rule[-2mm]{0mm}{0mm}
\langle\exp\{\,\, i \sum_{x\in S_\cC} \Omega_x\,\,\}\rangle & \sim & e^{-T V_{solid}(R)}\,,\\
\langle\exp\{\,\, i \sum_{x\in\cC} \alpha_x\,\,\}\rangle  & \sim & e^{-T V_{perim}(R)}\,,
\eea
where we have restricted ourselves to the consideration of rectangular $T\times R$,
$T \gg R$ loops only. Notice that $T, R$ dependence of Eq.~(\ref{exp})
is an ad hoc assumption which has to be checked separately. However, we have found
that (\ref{exp}) indeed accurately describe numerical data.
The details of simulations may be found in Ref.~\cite{self}. The final
result is shown on Fig.~2, where the solid curves are drawn to guide the eye.

There are few striking features of the expectation values (\ref{exp})
to be mentioned here. First of all, the 'perimeter' potential
turns out to be practically $R$-independent:
\beq
V_{perim}(R) ~\approx~  const\,,
\eeq
which might indicate that the perimeter contribution drops out in
Wilson loop VEV. Secondly, both $V_{dyn}(R)$
and $V_{solid}(R)$ are rising linear albeit with larger slope
than the full potential
\bea{c}
\label{linear}
\sigma_{dyn} \approx \sigma_{dyn} \approx 1.6 \, \sigma_{SU(2)}\,.
\eea
\centerline{\psfig{file=graph2.eps,width=0.45\textwidth,silent=,clip=}}
\centerline{Fig.~2}

~

\noindent
Moreover, $V_{solid}(R)$ is strictly linear starting from the smallest possible
distance $R=2$. The existence of a linearly rising term in the heavy quark potential at short
distances has been widely discussed in the literature, see, e.g.,
Refs.~\cite{Gubarev:1999ie} and references therein.

Finally, we emphasize that there is no factorization for Wilson loop VEV:
\beq
V(R) \ne V_{dyn}(R) + V_{solid}(R) + V_{perim}(R)\,.
\eeq
However the linear term $V_{solid}$ might survive at small distances since
the 'solid angle' contribution (\ref{exp}) is formally not suppressed by the
action even at very large $\beta$.

\end{document}